\documentclass[epj]{svjour}

\usepackage{graphics}
\usepackage{physics}
\usepackage{amssymb}
\usepackage{amsmath}

\begin{document}
\title{Building instructions for a ferromagnetic axion haloscope}
\author{Nicolò Crescini\inst{1} } 
\institute{Univ. Grenoble Alpes, CNRS, Grenoble INP, Institut Nel, 38000 Grenoble, France}

\date{Received: date / Revised version: date}

\abstract{
A ferromagnetic haloscope is a rf spin-magnetometer used for searching Dark Matter in the form of axions. A magnetic material is monitored searching for anomalous magnetization oscillations which can be induced by dark matter axions.
To properly devise such instrument one first needs to understand the features of the searched-for signal, namely the effective rf field of dark matter axions $B_a$ acting on electronic spins. Once the properties of $B_a$ are defined, the design and test of the apparatus may start. The optimal sample is a narrow linewidth and high spin-density material such as Yttrium Iron Garnet (YIG), coupled to a microwave cavity with almost matched linewidth to collect the signal. The power in the resonator is collected with an antenna and amplified with a Josephson Parametric amplifier, a quantum-limited device which, however, adds most of the setup noise. The signal is further amplified with low noise HEMT and down-converted for storage with an heterodyne receiver.
This work describes how to build such apparatus, with all the experimental details, the main issues one might face, and some solutions.
\PACS{
      {PACS-key}{discribing text of that key}   \and
      {PACS-key}{discribing text of that key}
     }
}

\maketitle

\section{Introduction}
\label{intro}
The axion is an hypothetical beyond the Standard Model particle, first introduced in the seventies as a consequence of the strong CP problem of QCD. Axions can be the main constituents of the galactic Dark Matter halos. Their experimental search can be carried out with Earth-based instruments immersed in the Milky Way's halo, which are therefore called ``haloscopes''. Nowadays haloscopes rely on the inverse Primakoff effect to detect  axion-induced excesses of photons in a microwave cavity under a static magnetic field.
This work describes the process leading to the successful operation of a ferromagnetic axion haloscope, which does not exploit the axion-to-photon conversion but its interaction with the electron spin.
The study of the axion-spin interaction and of the Dark Matter halo properties yields the features of the axionic signal, and is fundamental to devise a proper detector. A scheme of a realistic ferromagnetic haloscope is drawn to realize the challenges of its development.
It emerges that there are a number of requirements for a this setup to get to the sensitivity needed for a QCD-axion search. These are kept in mind when designing the prototypes, to overcome the problems without compromising other requirements.
A state-of-the-art sensitivity to rf signals allows for the detection of extremely weak signals as the axionic one. The number of monitored spins is necessarily large to increase the exposure of the setup, thus its scalability is a key part of the design process.

A ferromagnetic haloscope consists in a transducer of the axionic signal, which is then measured by a suitable detector.
The transducer is a hybrid system formed by a magnetic material coupled to a microwave cavity through a static magnetic field. Its two parts are separately studied to find the materials which match the detection conditions imposed by the axion-signal.
The detector is an amplifier, an HEMT or a JPA, reading out the power from the hybrid system collected by an antenna coupled to the cavity. A particular attention is given to the measurement of the noise temperature of the amplifier.
As it measures variation in the magnetization of the sample, the ferromagnetic haloscope is configured as a spin-magnetometer.

The present haloscope prototype \cite{PhysRevLett.124.171801} works at 90\,mK and reaches the sensitivity limit imposed by quantum mechanics, the Standard Quantum Limit, and can be improved only by quantum technologies like single photon counters.
The haloscope embodies a large quantity of magnetic material, i.\,e. ten 2\,mm YIG spheres, and is designed to be further up-scaled.  
This experimental apparatus meets the expected performances, and, to present knowledge, is the most sensitive rf spin-magnetometer existing. The minimum detectable field at 10.3\,GHz results in $5.5\times10^{-19}\,$T for 8\,h integration, and corresponds to a limit on the axion-electron coupling constant $g_{aee}\le1.7\times10^{-11}$. This result is the best limit on the DM-axions coupling to electron spins in a frequency span of about 150\,MHz, corresponding to an axion mass range from $42.4\,\mu$eV to $43.1\,\mu$eV.

The efforts to enhance the haloscope sensitivity include improvements in both the hybrid system and the detector.
The deposited axion power can be increased by means of a larger material volume, possibly with a narrower linewidth. To overcome the standard quantum limit of linear amplifiers one must rely on quantum counters. Novel studies on microwave photon counters, together with some preliminary results, are reported. Other possible usages of the spin-magnetometer are eventually discussed.

\section{Overview on axions}
\label{sec:theory}
A long-standing puzzle of beyond the Standard Model physics consists in the dark matter (DM) problem.
In 1933 Fritz Zwicky used two different techniques to estimate the mass of the Coma and Virgo clusters, one was based on the luminosity of the galaxies in the clusters, while the other used the velocity dispersion of individual galaxies. These two independent estimations did not agree by orders of magnitude \cite{zwicky}.
It is only in the seventies that this discrepancy started to be studied systematically. In particular, Vera Rubin studied the rotation curves of spiral galaxies and observed a violation of the second Kepler's law which can be explained assuming that the mass profile does not vanish beyond the stars \cite{rubin78,rubin80}. This was an early indication that spiral galaxies could be surrounded by an halo of DM.
Despite these evidences, the nature of DM is still unknown. Its possible composition could be baryonic or non-baryonic. The former considers matter similar to the one already known, while the latter comprehends hypothetical particles of beyond the Standard Model (BSM) physics.

The case of a non-baryonic DM is where cosmology meets particle physics. Approaching this problem, physicists glimpse the possibility of merging different questions which are apparently uncorrelated. New theories, remarkably supersymmetric DM \cite{susywimp}, triggered experimental searches in different forms and with various techniques. Low-background laboratory experiments aim at a direct detection \cite{wimprev}, accelerators could produce such particles and observe their missing energy and momentum \cite{Mitsou_2015}, while indirect evidences are based on their decay or annihilation \cite{FORNENGO20082010}. 
The detection of BSM particles would shed light on fundamental question like DM or the unification of all forces. Up to now the results of the LHC and experiments therein showed no evidence of new physics up to the 10\,TeV scale.
On the other hand, there are significant hints for physics at the sub-eV scale, like neutrino oscillation or the vacuum energy density of the Universe \cite{ringwald}. The physics case of weakly interacting sub-eV particles (WISPs) is motivated by the fact that any theory introducing a high-energy global symmetry breaking implies a light particle by the Nambu-Goldstone theorem \cite{weinbergpgb}.

Among other WISPs, the axion appears as a well-motivated BSM particle. Originally introduced to account for a fine-tuning issue in the SM known as the ``strong CP problem'' of quantum chromodynamics (QCD), it quickly became a prominent DM candidate.
The existence of axions is a very attractive perspective, since its addition to the Standard Model would solve two major problems of modern physics in a single shot \cite{Peccei2008}.
QCD is a non-Abelian $\mathrm{SU}(3)_c$ gauge theory which describes the strong interactions. Its Lagrangian ${\cal L}_\mathrm{QCD}$ contains a CP-violating term which is compatible with all symmetries of the SM gauge group.
However, since there is no experimental sign of CP violation in strong interactions, one needs to unnaturally suppress of this term.
The first $\mathrm{SU}(3)_c$ theory was proposed as CP-conserving to agree with experimental observations, but had an issue at low energy known as Weinberg's $\mathrm{U}(1)_A$ missing meson problem. The strong CP problem arises following the solution of the missing meson problem proposed by t'Hooft \cite{PhysRevLett.37.8,PhysRevD.14.3432}. This solution brings on CP violation in QCD, parametrized by $\bar{\theta}=\theta_\mathrm{QCD}+2\theta_Y$, two angles relative to QCD which, according to the theory, are independent. However, to conserve CP  either $\bar{\theta}$ should be zero or one of the quarks should be massless.

Among the measurable observables containing $\bar{\theta}$ and thus CP violation, the neutron electric dipole moment results \cite{tagkey1980487,POSPELOV2005119,doi:10.1146/annurev.ns.32.120182.001235}. Recent results by the nEDM collaboration constrain the parameter even more $\bar{\theta}\le10^{-10}$ \cite{PhysRev.108.120,Baker:2006ts,PhysRevD.92.092003}.
Since it is unlikely that nature chose very small $\theta_Y$ and $\theta_\mathrm{QCD}$ or a fine tuning among them, a naturalness problem arises. It is normally denoted as the strong CP problem, which is an important hint of BSM physics.
A solution to this problem is a scenario where $\bar{\theta}$ is promoted from a parameter to an actual particle.
This was realized, albeit in a different way, by Peccei and Quinn \cite{pq}, who introduced a new $\mathrm{U}(1)_\mathrm{PQ}$ symmetry to the SM to dynamically interpret $\bar{\theta}$. The idea was further developed by, among others, Weinberg and Wilczek \cite{wilczek1978problem,weinberg1978new,PhysRevLett.43.103,kim,SHIFMAN1980493}, who realized that it implies the existence of a new light pseudo-Nambu Goldstone boson which was called axion.
The minimization of the meson potential adjusts the axion vacuum expectation value to cancel any effect of CP violation, addressing the strong CP problem.

The axion mass eigenstate can then be computed from the masses of the pion $m_\pi$, of the up and down quarks, $m_u$ and $m_d$, and from the decay constants of the pion and of the axion itself, $f_\pi$ and $f_a$, resulting in the axion mass
\begin{equation}
m_a^2 \simeq = \frac{m_um_d}{(m_u+m_d)^2}\frac{m_\pi^2 f_\pi^2}{f_a^2} .
\label{eq:1_axmass}
\end{equation}
The energy scale $f_a$ is the PQ-symmetry breaking scale, and as the axion is the pseudo-Goldstone boson arising from this process, its mass and couplings are proportional to $f_a^{-1}$ making it very light and weakly-interacting.
The so-called ``invisible-axion models'' consider $f_a\simeq10^{12}\,\mathrm{GeV}$, and evade current experimental limits \cite{pdg}.
There are two main classes of invisible axions whose archetype are the Kim-Shifman-Vainshtein-Zakharov (KSVZ) and Dine-Fischler-Srednicki-Zhitnitsky (DFSZ) models \cite{Zhitnitsky:1980he,DINE1981199,SHIFMAN1980493,PhysRevLett.43.103,DINE1983137}, where the main difference is the coupling to SM particles.

It is now possible to analyze the axion as a constituent of DM.
Cold DM particles must be present in the Universe in a sufficient quantity to account for the observed DM abundance and they have to be effectively collisionless, i.\,e. to have only significant long-range gravitational interactions.
The axion satisfies both these criteria. Even if it is very light, the axion population is non-relativistic since it is produced out of equilibrium by vacuum realignment, string decay or domain wall decay \cite{abbott1983cosmological,PRESKILL1983127,DINE1983137,PhysRevD.32.3172,DAVIS1986225,PhysRevD.59.023505,LYTH1992279,PhysRevD.50.4821}. 
Being the main cold axions production mechanism, vacuum realignment is basically explained hereafter. In a time when the Universe cools down to a temperature lower than the axion mass, the axion field is sitting in a random point of its potential, and not necessarily at its minimum. As a consequence the field starts to oscillate and, since the axion has extremely weak couplings, it has no way to dissipate its energy. This relic energy density is a form of cold Dark Matter  \cite{MARSH20161,RevModPhys93015004}. 
Lattice QCD calculation can be used together with the present Dark Matter density to give an estimation of the QCD axion mass
\cite{bonati,BURGER2017880,berkowitz2015lattice,borsanyi2016calculation,petreczky2016topological,diCortona2016}

The axion has to be framed in the context of present physical theories, since one can wonder if the presence of light scalars may influence the behavior of already studied physical systems. Several constraints come from fitting the axion theory into astrophysical and cosmological observations. As other weakly interacting low-mass particles, they can contribute to the cooling of stars and be produced in astrophysical plasmas and in the Sun, contribute to stellar evolution and even affect supernovae \cite{raffelt1996stars,RAFFELT19901,turner1990windows,PhysRevD.79.107301,SCHLATTL1999353,redondo_2013,PhysRevLett.111.231301,RAFFELT1986402,C_rsico_2016,PhysRevLett.65.960,Leinson_2014,KELLER201362,PhysRevD.93.065044}. To sum up, these observations suggest that $m_a\le10\,$meV.
Cosmology provides both upped and lower limits for the axion mass, but being the upper ones weaker than the ones already described, the focus will be on lower bounds. 
These limits on the axion mass come from the production of DM-axions in the early Universe \cite{abbott1983cosmological,fox2004probing,Bae_2008,PhysRevD.82.123508,PhysRevD.74.123507,PhysRevD.75.103507,PhysRevD.78.083507,Hamann_2009,Planck2013,Planck2015}. In particular the axion mass must be higher than $6\,\mu$eV to avoid the overclosure problem, i.\,e. an axion density exceeding the observed DM density. Lighter masses are still possible within the so-called ``anthropic axion window''.

A general case of BSM particles are the so-called ``axion-like particles'' (ALPs). The interest in ALPs relies on the fact that its mass and coupling constants can be unrelated (other than for axions), thus they do not necessarily solve the strong-CP problem but still can account for the whole DM density of the Universe \cite{Arias_2012}. Any experimental search not reaching the axion sensitivity is still a probe of ALPs.

\subsection{Experimental searches}
\label{sec:expax}
In the last decades several experimental techniques have been proposed to detect axions and ALPs \cite{ringwald,redondo}.
Most experiments do not reach the axion-required sensitivity, but the physics result of these measurements is to limit the ALPs parameter space.
The most tested effects of axions are related to their coupling to photons, being this one the strongest and thus most accessible parameter. These limits mostly rely on the inverse Primakoff effect: in a strong static magnetic field it is possible to convert an itinerant axion into a photon that can be detected. Amongst all the experiment proposed or realised to detect axions, only haloscopes are treated in some details.

As already discussed axions may constitute DM, and if existing at least a fraction of DM have to be composed of axions. DM is an interesting source of axions and triggered multiple experimental searches.
Instruments searching for DM-axions composing the Milky Way's halo are called haloscopes. Haloscopes are particularly interesting in the scope of this work, which is devoted to the study of a ferromagnetic one.
In 1983, Sikivie proposed new ways to detect the axion by resonantly converting them into microwave photons inside a high quality factor ($Q$) cavity under a static magnetic field \cite{PhysRevLett.51.1415}. The resonance condition implies that the apparatus is sensitive to axions in a very narrow frequency range. The frequency of the axion signal is related to its mass and its width depends on the virial DM velocities in the galaxy. These kinds of experiments need to change resonant frequency to scan for different masses. The Axion Dark Matter eXperiment (ADMX) reached the sensitivity of KSVZ axions in the range $1.9\,\mu\mathrm{eV}-3.3\,\mu\mathrm{eV}$ \cite{PhysRevD.69.011101,PhysRevD.74.012006} assuming virialized axions composing the whole DM density $\varrho_\mathrm{DM}=0.45\,\mathrm{GeV}/\mathrm{cm}^3$. The setup was improved by using SQUID amplifiers \cite{PhysRevLett.104.041301}, and then reached the line of the DFSZ model \cite{PhysRevLett.120.151301,PhysRevD.103.032002}. The HAYSTAC experiment searched for heavier axions by operating a setup similar to the ADMX one but using a Josephson Parametric Amplifier (JPA), and achieving quantum limited sensitivity \cite{PhysRevLett.118.061302} and beyond \cite{backes_quantum_2021}. The collaborations UF and RBF also reached remarkable limits, and the ORGAN experiment operated a pathfinding haloscope at 110\,$\mu$eV \cite{PhysRevLett.59.839,MCALLISTER201767}.
Several new concepts have been proposed to search for DM axions with next-generation haloscopes based not only on the Primakoff effect but also on axion-induced electric dipole moments or on the axion-spin interaction \cite{capp,caldwell2017dielectric,PhysRevX.4.021030,Garcon_2017,1742-6596-718-4-042051,BARBIERI2017135,PhysRevLett.122.121802,Caspers:1989ix}.

\section{The effective magnetic field of DM axions}
\label{sec:signal}
The coupling between axions and electron spins can be used for axion detection as an alternative to the coupling to photons \cite{BARBIERI1989357,kakhidze1991antiferromagnetic,vorobyov1995ferromagnetic,PhysRevLett.55.1797}. Being it weaker than the axion-photon coupling it was not immediately exploited, but recently new experimental schemes were presented. Besides the axion discovery, the axion-electron coupling is interesting for distinguishing between different axion models.
The possibility of detecting galactic axions by means of converting them into collective excitations of the magnetization (magnons) was considered by Barbieri \textit{et al.} in \cite{BARBIERI1989357}, laying the foundations to the Barbieri Cerdonio Fiorentini Vitale (BCFV) scheme and to the following experimental proposal \cite{BARBIERI2017135}.
The original idea is to use the large de Broglie wavelength of the galactic axions to detect the coherent interaction between the axion DM cloud and the homogeneous magnetization of a macroscopic sample.
To couple a single magnetization mode to the axion field, the sample is inserted in a static magnetic field.
The interaction yields a conversion rate of axions to magnons which can be measured by monitoring the power spectrum of the magnetization.
The form of the interaction is calculated hereafter in terms of an effective magnetic field. Such a field is the searched-for signal. Its features are derived and characterized as follows.

\begin{figure}[h!]
	\centering
	\resizebox{0.2\hsize}{!}{ \includegraphics*{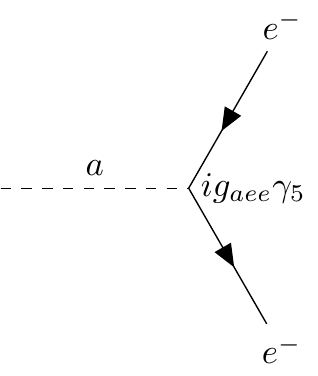}} 
	\caption{Feynman diagrams of the axion-fermion interaction, showing how the effect of the axion is to be absorbed and cause the spin flip of the fermion. The corresponding interaction Lagrangian is reported in Eq.\,(\ref{eq:2_lagaf}).}
	\label{fig:fey}
	\end{figure}

The axion derivative interaction with fermions is invariant under a shift of the axion field $a\rightarrow a+a_0$ and reads
\begin{equation}
{\cal L}_\psi = \frac{C_\psi}{2f_a} \bar{\psi} \gamma^\mu \gamma_5 \psi \partial_\mu a
\label{eq:2_lagaf1}
\end{equation}
where $\psi$ is the spinor field of a fermion of mass $m_\psi$, and $C_\psi$ is a model-dependent coefficient.
The dimensionless couplings can be defined as
\begin{equation}
g_{a\psi\psi}=C_\psi m_\psi/f_a
\label{eq:2_dlaxc}
\end{equation}
and play the role of Yukawa couplings, while the fine structure constant of the interaction is $\alpha_{a\psi\psi}=g^2_{a\psi\psi}/4\pi$.
The tree-level coupling coefficient to the electrons of the DFSZ model is \cite{Zhitnitsky:1980he,DINE1981199} $C_e = \cos^2 \beta^\prime / 3$, where $\tan \beta^\prime = v_d/v_u$, the ratio of the vacuum expectation values of the Higgs field.
The axion-electron derivative part of the interaction can be expressed as
\begin{equation}
{\cal L}_e = \frac{g_{aee}}{2m_e}\partial_\mu a(x) \big(\bar{e}(x)\gamma^\mu\gamma_5 e(x)\big) \simeq -ig_{aee}  a(x) \bar{e}(x) \gamma_5 e(x),
\label{eq:2_lagaf}
\end{equation}
where the last term is an equivalent Lagrangian obtained by using Dirac equation and neglecting quadridivergences. The Feynman diagram of this interaction is reported in Fig.\,\ref{fig:fey} and suggests how the process happens: an axion is absorbed and causes the fermion to flip its spin, and the macroscopic effect is a change in the magnetization of the sample containing the spin.
	
By taking the non-relativistic limit of the Euler–Lagrange equation, the time evolution of a spin 1/2 particle can be described by the usual Schroedinger equation
\begin{equation}
i\hbar \pdv{\varphi}{t}=\Big( -\frac{\hbar^2}{2 m_e} \nabla^2 -\frac{g_{aee} \hbar}{2m_e} \boldsymbol{\sigma}_e\cdot \nabla a  \Big)\varphi,
\label{eq:2_em_nrl}
\end{equation}
where $\sigma_e$ is the Pauli matrices spin vector.
The first term on the right side of Eq.\,(\ref{eq:2_em_nrl}) is the usual kinetic energy of the particle, while the second one is analogous to the interaction between a spin and a magnetic field.
One can notice that
\begin{equation}
 -\frac{g_{aee} \hbar}{2m_e} \boldsymbol{\sigma}_e\cdot \nabla a  = -2\frac{e\hbar}{2m_e} \boldsymbol{\sigma}_e \cdot \Big( \frac{g_{aee}}{2e} \Big)\nabla a = -2\mu_e \boldsymbol{\sigma}_e \cdot \Big( \frac{g_{aee}}{2e} \Big)\nabla a,
\label{eq:2_int_nrl}
\end{equation}
since $\mu_e$ is the magnetic moment of the particle, it can be both Bohr magneton or a nuclear magneton, depending on the considered fermion.
From Eq.\,(\ref{eq:2_int_nrl}) it is clear that the effect of the axion is the one of a magnetic field, but since it does not respect Maxwell's equations, calling it an effective magnetic field is more appropriate
\begin{equation}
\mathbf{B}_a \equiv \Big( \frac{g_{aee}}{2e} \Big)\nabla a.
\label{eq:2_Ba}
\end{equation}
This definition is useful to quantify the performances of a ferromagnetic haloscope in terms of usual magnetometers sensitivity.
In the BCFV case, the signal is given by the electrons’ magnetization. An intuitive connection between  the macroscopic magnetization and the spin is given by $M\propto\mu_B N_S$ where $\mu_B$ is Bohr magneton and $N_S$ is the number of spins which take part to the magnetic mode \cite{landau1980statistical}. In such a way, the spin-flip can be classically considered a variation of the magnetization at a frequency given by the axion field.

It is now necessary to understand which are the features of $B_a$ to design a proper detector. The isothermal model of the Milky Way's DM halo predict a local density of $\varrho_\mathrm{DM}\simeq 0.45\,\mathrm{GeV}/\mathrm{cm}^3$ \cite{cmbdm1}. An Earth-based laboratory is thus subjected to an axion-wind with a speed $v_a\simeq300\,$km/s, that is the relative speed of Earth through the Milky Way.
Using the vector notation, the value of $v_a^i$ can be calculated from the speed of the galactic rest frame. The speed on Earth $v_E^i$ is given by the sum of  $v_S^i$, $v_O^i$ and $v^i_R$, which are respectively the Sun velocity in the galactic rest frame (magnitude 230\,km/s), the Earth’s orbital velocity around the Sun (magnitude 29.8\,km/s), and  the Earth’s rotational velocity (magnitude 0.46\,km/s). The observed axion velocity is then $v_a^i = - v_E^i$, which follow a Maxwell-Boltzmann distribution.
As will be shown hereafter, the effect of this motion is a non-zero value of the axion gradient, and a modulation of the signal with a periodicity of one sidereal day and one sidereal year \cite{turner1990windows,PhysRevD.42.3572,PhysRevD.42.1001}.

The numeric axion density in the DM halo depends on the axion mass and results $n_a\simeq3\times10^{12}\,(10^{-4}\,\mathrm{eV}/m_a)\,\mathrm{cm}^{-3}$. The coherence length of the axion field is related to the de Broglie wavelength of the particles, which is given by
\begin{equation}
\lambda_a=\frac{h}{m_av_a}\simeq14\,\Big( \frac{10^{-4}\,\mathrm{eV}}{m_a} \Big)\,\mathrm{m}.
\label{eq:2_lambdaa}
\end{equation}
Such wavelength allows for the use of macroscopic samples to detect the variation of the magnetization.
The large occupation number $n_a$, coherence length $\lambda_a$, and $\beta_a=v_a/c\simeq10^{-3}$ permit to treat $B_a$ as a classical field\footnote{The average speed $v_a\ll c$ also justifies the approximation of Eq.\,(\ref{eq:2_em_nrl}), i.\,e. the use of the non-relativistic limit of Euler-Lagrange equations.}. The coherent interaction of $a(x)$ with fermions has a mean value
\begin{equation}
a(x) = a_0 e^{i p_a^\mu x_\mu} = a_0 e^{i(p_a^0 t - p_a^i x_i)},
\label{eq:2_axionfield}
\end{equation}
where $p_a^i=m_a v_E^i$ and $p_a^0=\sqrt{m_a^2+|p_a^i|^2}\simeq m_a+|p_a^i|^2/(2m_a)$.
The production of DM axions is discussed in Section \ref{sec:theory}, where it is shown that they are indeed cold DM since their momentum is orders of magnitude smaller than the mass.
The axion kinetic energy is expected to be distributed according to a Maxwell-Boltzmann distribution, with a mean relative to the rest mass of $7\times10^{-7}$ and a dispersion about the mean of $\sigma_\mathrm{MB}\simeq5\times10^{-7}$ \cite{turner1990windows,PhysRevD.42.3572,0004-637X-845-2-121}.
The effect of the mean is a negligible shift of the resonance frequency with respect to the axion mass. The consequence of the dispersion on the effective magnetic field is a natural figure of merit
\begin{equation}
Q_a=\frac{1}{\sigma_\mathrm{MB}}\simeq\Big(\frac{m_a}{\langle p^i_a\rangle}\Big)^2=\frac{1}{\beta_a^2}\simeq2\times10^6.
\label{eq:2_axq}
\end{equation}
To calculate the field amplitude $a_0$, the momentum density of the axion field is equated to the mean DM momentum density yielding
\begin{equation}
a_0^2p_1^0p_a^i = n_a\langle p^i_a \rangle = n_a m_a v_a \quad \Rightarrow \quad a_0=\sqrt{n_a/m_a}.
\label{eq:2_a0calc}
\end{equation}
For calculation purposes natural units are dropped and the Planck constant $\hbar$ and speed of light $c$ are restored. The effective magnetic field associated to the mean axion field reads
\begin{equation}
B_a^i = \frac{g_{aee}}{2e}\Big( \frac{n_a \hbar}{m_a c} \Big)^{1/2} p_a^i \sin\Big( \frac{p_a^0 ct + p_a^i x_i}{\hbar} \Big).
\label{eq:2_baxeff}
\end{equation}
From Eq.\,(\ref{eq:2_baxeff}), the frequency and amplitude of the axionic field interacting with electrons result
\label{eq:2_axeffb}
\begin{align}
\begin{split}
B_a =& \frac{g_{aee}}{2e}\Big( \frac{n_a \hbar}{m_a c} \Big)^{1/2} m_a v_a \simeq 5\times10^{-23}\, \Big( \frac{m_a}{50\,\mu\mathrm{eV}} \Big)\,\mathrm{T},    \\
\frac{\omega_a}{2\pi} \simeq& \,\frac{cp_a^0}{\hbar} = \frac{m_ac^2}{\hbar} \simeq 12\,\Big( \frac{m_a}{50\,\mu\mathrm{eV}} \Big)\,\mathrm{GHz}.
\label{eq:2_axpar1}
\end{split}
\end{align}
As the equivalent magnetic field is not directly associated to the axion field but to its gradient, the corresponding correlation length and coherence time must be corrected to \cite{BARBIERI2017135}
\begin{align}
\begin{split}
\lambda_{\nabla a} \simeq& \,0.74\, \lambda_a = 0.74\frac{\hbar}{m_a v_a} \simeq 20\, \Big( \frac{50\,\mu\mathrm{eV}}{m_a} \Big)\,\mathrm{m};    \\
\tau_{\nabla a} \simeq& \,0.68\,\tau_a = 0.68\frac{2\pi \hbar}{m_av_a^2} \simeq 46\, \Big( \frac{50\,\mu\mathrm{eV}}{m_a} \Big)\Big( \frac{Q_a}{1.9\times10^6} \Big)\,\mu\mathrm{s}.
\label{eq:2_axpar2}
\end{split}
\end{align}
The nature of the DM axion signal is now well-defined: an effective magnetic field of amplitude $B_a$, frequency $f_a=\omega_a/2\pi$, and quality factor $Q_a$ with values defined by Eq.s\,(\ref{eq:2_axpar1}) and (\ref{eq:2_axpar2}).

\section{The axion-to-electromagnetic field transducer}
\label{sec:transducer}
A viable experimental scheme must be designed to detect the field $B_a^i$, whose features are defined by Eq.s\,(\ref{eq:2_axpar1}) and (\ref{eq:2_axpar2}). A magnetic sample with a high spin density $n_S$ and a narrow linewidth $\gamma_m=2\pi/T_2$ (i.\,e. long spin-spin relaxation time $T_2$) can be used as a detector.
The magnetic field $B_a$ drives a coherent oscillation of the magnetization over a maximum volume of scale $\lambda_{\nabla a}$. The sensitivity increases with the sample volume $V_s$ up to $(\lambda_{\nabla a})^3$.

For electrons $\gamma_e\simeq(2\pi)28\,$GHz/T, so the corresponding magnetic field $B_0$ is of order 1\,T and experimentally readily obtainable.
The electrons' spins of a magnetic sample under a uniform and constant magnetic field result in a magnetization $\mathbf{M}(\mathbf{x},t)$ that can be divided in magnetostatic modes. The space-independent mode of uniform precession is called Kittel mode.
The axionic field couples to the components of $\mathbf{M}$ transverse to the external field, depositing power in the material. More power is deposited if the axion field is coherent with the Kittel mode for a longer time.
The best-case scenario is a material with a quality factor $Q_m= \gamma_e B_0/\gamma_m$ which matches $Q_a$, so that the coherent interaction between spins and DM-axions lasts for $\tau_{\nabla a}$. For this reason the magnetic field uniformity over the sample must be $\le 1/Q_m$ to avoid inhomogeneous broadening of the ESR.

\begin{figure}[h!]
	\centering
	\resizebox{0.5\hsize}{!}{\includegraphics{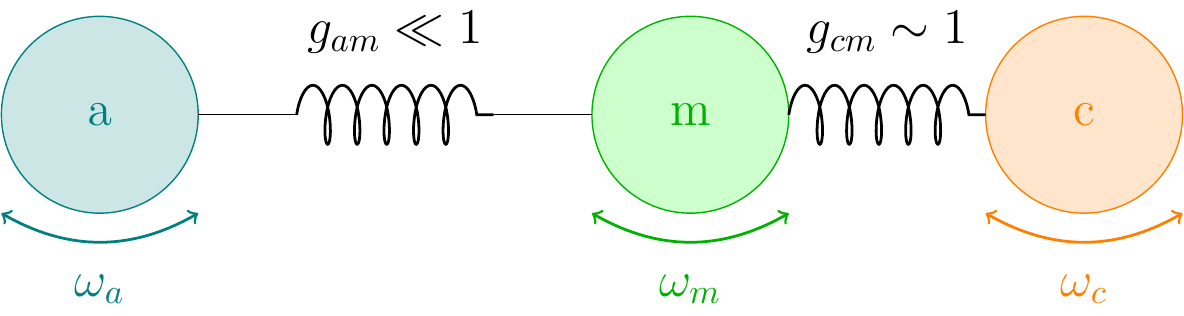}}
	\caption{The coupled harmonic oscillators are reported in orange, green and blue for cavity $c$, material $m$ and axion $a$ respectively. The uncoupled normal-modes frequencies of the HOs are $\omega_c$, $\omega_m$ and $\omega_a$ and the couplings are $g_{am}$ and $g_{cm}$, represented by springs.}
	\label{fig:ho}
	\end{figure}

% quindi misuriamo l'assione solo prendendo un pezzo di materiale e vedendo la magnetizzazione
According to these considerations, it is possible to detect an axion-induced oscillation of the magnetisation by monitoring a large sample with an precise magnetometer.
However, the limit of this scheme lies in the short coherence time of the magnetic sample. In fact at high frequency, i.\,e. above 1\,GHz, the rate of dipole emission becomes higher than the intrinsic material dissipation, this effect is know as radiation damping \cite{PhysRev.95.8}.
Since radiation damping is related to the sample dipole emission, a possible way to reduce its contribution is to limit the phase-space of the radiated light by working in a controlled environment like a resonant cavity \cite{PhysRev.95.8,kittel,kittel2,PhysRev.170.379}.  
By housing the sample in a mw cavity and tuning the static magnetic field such that $\omega_m=\gamma_eB_0\simeq\omega_c$, where $\omega_c$ is the resonance frequency of a cavity mode with linewidth $\gamma_c$, one obtains a photon-magnon hybrid system (PMHS).
An exact description of the system is given by the Tavis-Cummings model \cite{PhysRev.170.379}. It discusses the interaction of $N_S$ two-level systems with a single mw mode, and predicts a scaling of the cavity-material coupling strength $g_{cm}\propto\sqrt{N_S}$. The single-spin coupling is
    \begin{equation}
    g_s=\frac{\gamma_e}{2\pi}\sqrt{\frac{\mu_0 \hbar \omega_m}{\xi V_{c}}},
    \label{eq:3_gs_coupling}
    \end{equation}
where $V_c$ is the cavity volume, $\mu_0$ is the vacuum magnetic permeability and $\xi$ a mode-dependent form factor \cite{PhysRevLett.113.156401}, with this relation $g_{cm}=g_s\sqrt{N_S}$.
Such scaling has been verified experimentally down to mK temperatures for an increasing number of spins $N_S$ \cite{PhysRevLett.113.156401,PhysRevLett.113.083603,PhysRevB.104.064426}.
For a quantity of material such that $g_{cm}\gg \gamma_c$, the single cavity mode splits into two hybrid modes with frequencies $\omega_+$, $\omega_-$ and $2g_{cm}=\omega_+-\omega_-$.
For $\omega_m=\omega_c$ the linewidths of the hybrid modes are the average of the cavity mode linewidth $\gamma_c$ and of the material one $\gamma_m$, namely $\gamma_h=(\gamma_c+\gamma_m)/2$.
The coupling $g_{cm}$ is in fact a conversion rate of the material magnetization quanta (magnons) to cavity photons and viceversa. If $g_{cm}>\gamma_h$ the system is in the strong-coupling regime, meaning that for a magnon (photon) it is more likely to be converted than to be dissipated. In this way, magnetisation fluctuations, which might be induced by axions, are continuously converted to electromagnetic radiation that be collected with an antenna coupled to the cavity mode, as schematically shown in Fig.\,\ref{fig:ho}.

\begin{figure}[h!]
	\sidecaption
	\resizebox{0.4\hsize}{!}{ \includegraphics*{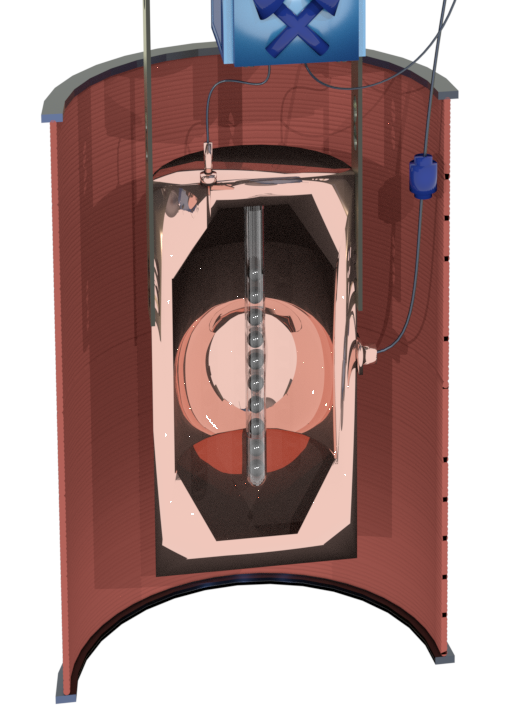}} 
	\caption{Rendering of the whole system, constituted by the cavity and the pipe with ten YIG spheres, ready to be tested at milli-Kelvin temperatures. The external part shows the superconducting magnet (in brown) which surrounds the cavity and the spheres to provide a magnetic field with uniformity better than 7\,ppm. The magnet is immersed in the liquid helium bath outside the vacuum chamber of the dilution unit. The cavity is at the centre of the magnet, is anchored to the mixing chamber of the dilution refrigerator with two copper bars and is equipped with two antennas, one is fixed and weakly coupled, while the second one is movable and is used to extract the signal. The YIG spheres are inside the cavity, held by a fused silica pipe filled with helium and separated by thin PTFE spacers. The cap used to seal the pipe is made of copper and is anchored to the cavity body to ensure the thermalisation of the exchange helium and therefore of the YIG spheres.}
	\label{fig:transducer}
	\end{figure}

The aim of a PMHS devised for an axion haloscope is to maximise the axionic signal, and it effectively works as an axion-to-photon transducer. A rendering of the resulting device is reported in Fig.\,\ref{fig:transducer}.

An important result for designing the transducer is that multiple spheres can be coherently coupled to a single cavity mode \cite{quaxepjc,PhysRevB.104.064426}. The measurements demonstrate that all the spins participate in the interaction, thus the samples act as a single oscillator.
This is guaranteed by the fact that the static field is uniform over the spheres and that the rf field is degenerate over the axis of the cavity where they are placed.
Several tests were performed to understand different features of the system in the light of the two properties mentioned before.
To understand the results of the different measurements one can use a simple oscillators model as is done in \cite{PhysRevB.104.064426}. 
The PMHS can be described by introducing two magnon modes and two cavity modes, hereafter the photon modes are labeled as $c$ and $d$ while the magnon modes are $m$ and $n$. In the matrix form, the system can be modeled by the hamiltonian
\begin{equation}
{\cal H}_\mathrm{cdmn} =
    \begin{pmatrix}
    \omega_c -i\gamma_c/2     &    g_{cd}    &    g_{cm}        &    g_{cn}    \\
    g_{cd}        &    \omega_d -i\gamma_d/2    &    g_{dm}    &    g_{dn}    \\
    g_{cm}         &    g_{dm}    &    \omega_m -i\gamma_m/2     &    g_{mn}    \\
    g_{cn}        &    g_{dn}    &    g_{mn}    &    \omega_ n -i\gamma_n/2     \\
    \end{pmatrix},
\label{eq:mncd_osc}
\end{equation}
where $\omega$, $\gamma$ and $g$ are the frequencies, linewidth and coupling of the different modes.
The autofunction of the system can be calculated as the determinant of $\omega\mathbb{I}_4 - {\cal H}_\mathrm{cdmn}$ thus the function used to show the anticrossing curve reads
\begin{equation}
f_\mathrm{cdmn}(\omega) = \det \big( \omega\mathbb{I}_4 - {\cal H}_\mathrm{cdmn} \big).
\label{eq:4_autofcmn}
\end{equation}

To ideally have a coherent coupling, one needs to let the spins of different spheres cooperate and make them indistinguishable, so they need to be uncoupled and their resonant frequencies must be the same.
These conditions translate to $g_{mn}=0$ and $\omega_m=\omega_n$, which clearly can be extended to an arbitrary number of oscillators (in this case, the ten spheres).
The interaction between two spheres yields non-zero value of $g_{mn}$, and its effect is to introduce other resonances besides the two main ones of the PMHS. This effect needs to be avoided to have control over the system and couple all the spins of the samples to the cavity mode, avoiding magnons bouncing between different magnetic modes and eventually being dissipated before their photon conversion.

YIG sphere were produced on site with a technique described in \cite{PhysRevB.104.064426}. This open the possibility of studying spheres of different diameters coupled to the same mode. One of the findings is that, trying to couple spheres with different diameter to the same mode, the volume of the sample is linearly related to the offset field \cite{PhysRevB.104.064426}.
The axion-to-electromagnetic field transducer of a ferromagnetic haloscope. The constraints to remember for its design are in the following, and were tested with a room temperature setup consisting in a 10.7\,GHz cavity with conical endcaps and a fused silica pipe holding the YIG spheres. The magnetic field is given by a SC magnet which, to perform quick tests, it is equipped with a room temperature bore allowing the magnet to be in a liquid helium bath during operation.
First the minimum separation between two spheres is tested by gradually increasing the distance between them and verifying that a usual anticrossing curve is reproduced. The minimum distance between 2\,mm spheres results in 3\,mm. A YIG sample then occupies 5\,mm of space, and since the cylindrical part of the cavity is 6\,cm it can house a maximum of twelve samples. Ten spheres are inserted in the pipe for them not to be too close to the conical part of the cavity. Multiple spheres of different diameters were fabricated and refined to verify that they hybridize with the cavity for the same value of the magnetic field.

The setup must ensure a proper thermalization of the cavity and of the YIG spheres, the preparation of the fused silica pipe is as follows.
A vacuum system is designed in such a way to empty the pipe from air which is then immersed in a 1\,bar helium controlled atmosphere. This way the pipe is filled with helium, and can be sealed by using a copper plug and Stycast.
First the sealing is tested without the samples by measuring the shift of the TM110 mode of the cavity-pipe system with and without helium. The frequency is measured with the helium-filled pipe, which is then immersed in liquid nitrogen and again placed in the cavity. Re-measuring the same frequency excludes the presence of leaks.

The used cavity is made of oxygen-free high-conductivity copper, and features a cylindrical body with two conical endcaps, as shown in Fig.\,\ref{fig:transducer}. The central body is not a perfect cylinder but it has two flat surfaces used to remove the angular degeneration of the mode. This creates two modes rotated of $\pi/2$ with different frequencies, which is the second cavity mode in Eq.\,(\ref{eq:mncd_osc}). The function $f_\mathrm{cdmn}(\omega)$ is fitted to the measured PMHS dispersion relation to extract the parameter of our setup,  and in particular the hybridization results 638\,MHz which is compatible with the single 1\,mm sphere since $638\,\mathrm{MHz}/\sqrt{8\times10}=71\,$MHz. This value of the single sphere coupling is compatible with what previously obtained in simpler PMHS, indicating that the measured spin density of YIG is consistent both with the previous results and with the values reported in the literature.
Remarkably, the lower frequency resonance is almost unaffected by the behaviour of the rest of the PMHS, in the sense that its frequency does not differ from the one of a usual anticrossing curve, thus it can possibly be safely used for a measurement \cite{rfband}.

Since haloscopes need to scan multiple frequencies to search for axions, the resonant frequency of the PMHS mode used for the measurement need to be changed. The tuning is made extremely easy by the fact that it is controlled only by means of the external magnetic field. A high stability of $B_0$ is necessary to perform long measurements over a single frequency band. This is set by the linewidth of the hybrid mode, which in this case is 2\,MHz, and is tuned to cover a range close to 100\,MHz \cite{PhysRevLett.124.171801,rfband}.
Thanks to the anticrossing curve it is easy to identify the frequency of the correct mode to study. The hybrid mode is not affected by disturbances caused by other modes in a range that largely exceeds ten times its linewidth. These clean frequencies are selected for the measurements whenever it is possible to match them with the working frequencies of the amplifier described in the next Section.

\section{Quantum-limited amplification chain}
\label{sec:amplifiers}
The PMHS described previously in this Chapter acts as a transducer of the axionic signal. The power coming from the PMHS must be measured and acquired with a suitable detection chain, and, as it is extremely weak, needs to be amplified. The intrinsic noise of an haloscope is essentially related to the temperature of the setup, and since axionic and Johnson power have the same origin it is the ultimate limit of the SNR. The amplification process inevitably introduces a technical noise which, for these setups, is useful to quantify in terms of noise temperature $T_n$ to compare to the Johnson noise.
This stage of the measurement is setting the overall sensitivity of the apparatus since, as shown hereafter, for very low working temperatures the noise temperature is higher than the thermodynamic one.
Minimizing $T_n$ is a key part of the development of an haloscope, and is complementary to the maximization of the axion deposited power.
The mw amplifiers used for precision measurement are mostly high electron mobility transistors (HEMT), since they have high gain and low noise, of the order of 4\,K. The most sensitive amplifier available is the Josephson parametric amplifier (JPA), which reaches the quantum noise limit \cite{PhysRevA.39.2519,jpacastellanos,doi:10.1063/1.2929367,doi:10.1063/1.2750520,doi:10.1063/1.2964182,Abdo_2009,doi:10.1063/1.4886408}. This type of amplifier is used in the present haloscope, and its performances can be overcome only by using a photon counter.

HEMT are field effect transistors based on an heterojunction, i.\,e. a PN junction of two materials with different band gaps \cite{989961}.
The proper doping profile and band alignment gives rise to extremely high electron mobilities, and thus to amplifiers which can have high gain, very low noise temperature, and working frequency in the microwave domain.
Even if their noise temperature is low, HEMTs are not the most sensitive amplifiers available.

\begin{figure}[h!]
	\sidecaption
	\resizebox{0.4\hsize}{!}{ \includegraphics*{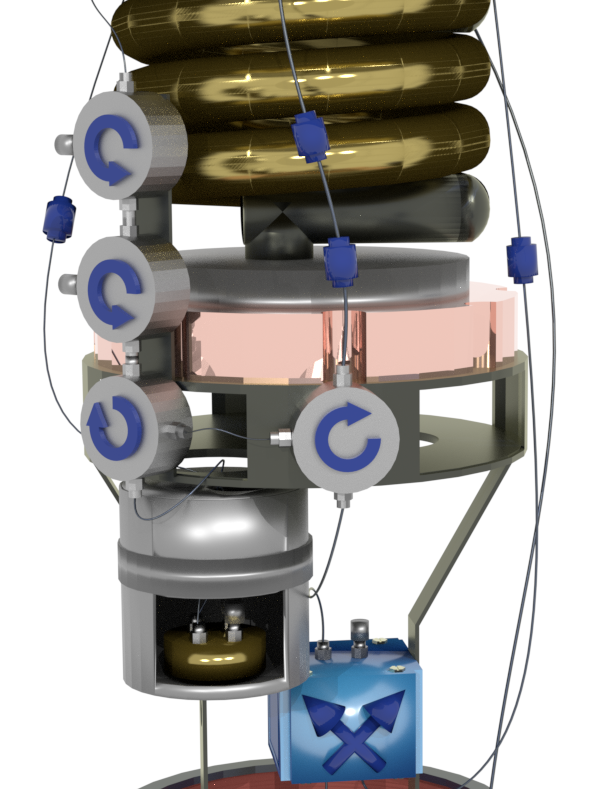}} 
	\caption{Rendering of the implementation of this JPA in a ferromagnetic haloscope. The golden pipe is connected to the mixing chamber of the dilution refrigerator used to cool down the setup, the circulators are only in thermal contact with this last stage, as is the shielding cage of the JPA (also drawn in gold). The blue component is a switch, present in one of the possible configurations of a ferromagnetic haloscope; attenuators are drawn in blue as well. The JPA is inside two concentric cans, the external one is made of Amuneal and the external is of aluminum. The first is useful to reduce the Earth magnetic field in which the superconducting parts (shields and junctions) undergo the transition, while the second screens from external disturbances. Everything is attached to the mixing chamber plate of a dilution refrigerator with a base temperature of about 90\,mK. This image corresponds to the configuration reported in Fig.\,\ref{fig:rf}b.}
	\label{fig:amplification}
	\end{figure}

At a frequency 10\,GHz the SQL of linear amplifiers is close to 0.5\,K, which is about one order of magnitude lower than the $T_n$ of HEMTs. Such remarkably low $T_n$ is achieved by JPAs, resonant amplifiers with a narrow bandwidth but with quantum-limited noise. This feature makes them the ideal tool to measure faint rf signals, and thus to be implemented in ferromagnetic or Primakoff haloscopes.
The non-linear mixing is given by a Josephson-RLC circuit with a quadratic time-dependent Hamiltonian, which can be degenerate or non-degenerate depending on whether the signal and idler waves are at the same frequency or not \cite{ROY2016740}.
A non-degenerate device consists in a three-modes, three-input circuit made of four Josephson junctions forming a Josephson ring modulator. It effectively is a three-wave purely dispersive mixer which can be used for parametric amplification \cite{Bergeal2010}. It can be computed that the non-linear mixing process appears as a linear scattering, configuring the JPA is a linear amplifier. As such, it is quantum limited and its noise temperature depends on the working frequency.
Being based on resonant phenomena, the JPA has a narrow working band of tens of MHz. 
To use the amplifier in a wide frequency range, a bias field is applied to the ring and the resonance frequencies of the signal, idler and pump mode are tuned. This is achieved with a small SC coil placed below the ring, biased with a current $I_b$. The implementation of a JPA in a ferromagnetic haloscope is shown and described in Fig.\,\ref{fig:amplification}.

\begin{figure}[h!]
	\centering
	\resizebox{0.36\hsize}{!}{ \includegraphics*{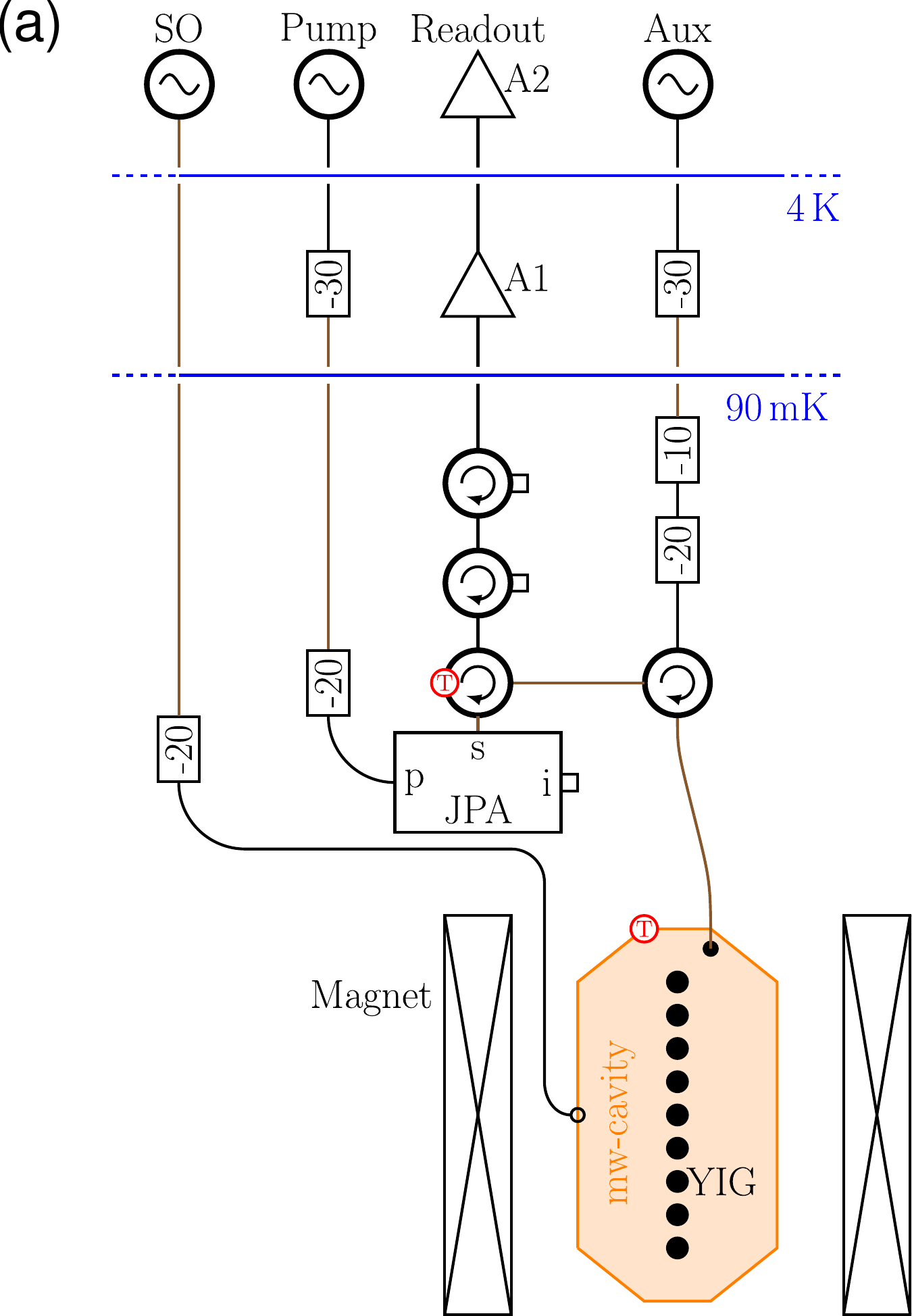}} 
	\resizebox{0.4\hsize}{!}{ \includegraphics*{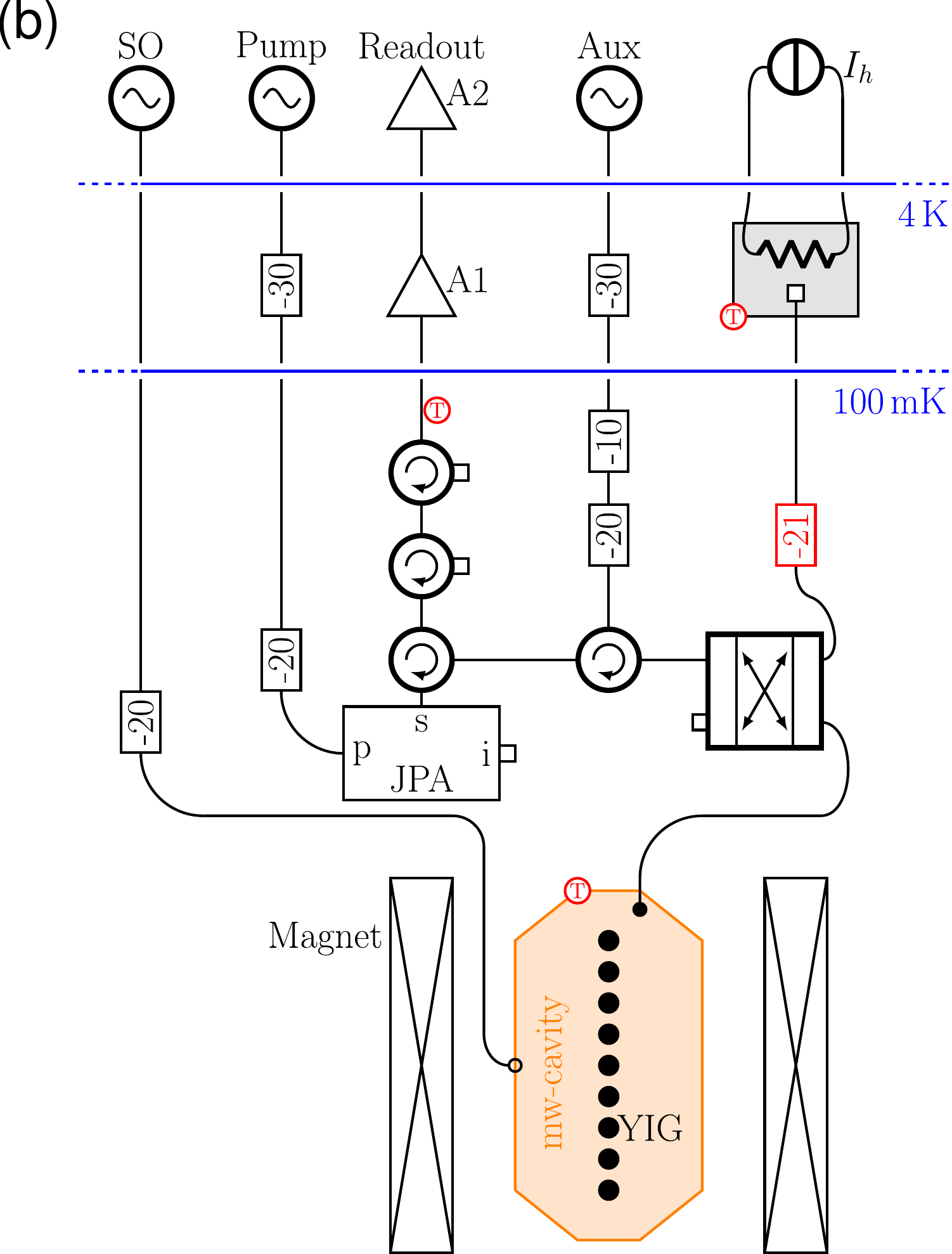}} 
	\caption{Two possible electronics layout for a ferromagnetic haloscope. The blue lines show the temperature ranges, the crossed rectangles are the magnet, and the orange rectangle is the cavity with black YIG circles inside. The boxed numbers are attenuators and the red circled $T$s are the thermometers. At the top of the cavity are located the weakly coupled antenna (empty dot) and the variably-coupled antenna (full dot). The weakly coupled antenna is connected to an attenuator and then to the source oscillator SO. In configuration (a) the the variable antenna is connected to the JPA through a circulator, whose other input is used for auxiliary measurements. The output of the JPA is further amplified by two HEMTs A1 and A2.
Configuration (b) is basically the same as (a), where the input can be switched from the cavity antenna to a matched load with variable temperature regulated by a current $I_h$, and used for calibration. In both (a) and (b) the A2 output is down-converted and acquired.}
	\label{fig:rf}
	\end{figure}

The characterisation the rf chain used for the measurements is described hereafter. It will focus on the setup described by Fig.\,\ref{fig:rf}a, as is the one used in \cite{PhysRevLett.124.171801} as it was found to be more reproducible and in general more reliable than \ref{fig:rf}b.
The configuration of the electronics allows the testing of both the JPA and the PMHS. Transmission measurements of the PMHS can be performed by turning off the JPA (i.e. no bias field and no pump) to reflect the signal on it, the input is the SO line and the output is the readout line. The JPA can be tested with the help of the Aux line, by uncoupling the antenna from the cavity and reflecting the incoming signal. Some rf is still absorbed at the cavity modes frequency but this does not compromise the measurement.
The external static field of the PMHS does not affect the resonances of the Josephson ring modulator as no difference has been detected between the measurements with and without field.
Runs are performed with bias currents $I_b\simeq170\,\mu$A and $460\,\mu$A at frequencies ranging from 10.26\,GHz to 10.42\,GHz.

Using the SO line and critically coupling the antenna to the hybrid mode, a signal is injected in the system and read with the whole amplification chain. The first test is to verify the linearity of the JPA (and of the whole chain) using signals of growing intensity until the system saturates. These measurements show the linear and saturate behavior of the amplifier. It is possible to calibrate the gain of the JPA by using a signal large enough to be measured with the JPA off but also not to saturate it once it is turned on.
This is useful to know the gain of the amplifier at the different working points to have a preliminary calibration of the system and to understand whether an output noise with higher amplitude is due to the JPA or to something else.
Since the electronics above the 4\,K line was already characterized for the previous prototype, the baseline noise with JPA off is roughly the amplifier noise temperature $T_n^\mathrm{(hemt)}\simeq10\,$K. The measured noise spectra with the JPA turned off is white in a bandwidth of several hundreds of MHz, when the parametric amplifier is turned on its resonance exceeds this noise of roughly 10\,dB. Since it is possible to calibrate the gain of the JPA $G_\mathrm{JPA}\simeq20\,$dB, the amplified noise level can be extracted as $T_n^\mathrm{(JPA)}=T_n^\mathrm{(hemt)}/10^{(G_\mathrm{JPA}-10\,\mathrm{dB})/10}\simeq 1\,$K, which is the noise temperature amplified by the JPA.
The value of 1\,K is reasonable, since a single quantum at this frequency is 0.5\,K such noise corresponds to two quanta. Even if this procedure is somewhat correct, it is not a proper calibration of the setup and something better is explained hereafter.

Since some problems were encountered in the noise calibration with hot load (see Fig.\,\ref{fig:rf}b), the rf setup of Fig.\,\ref{fig:rf}a is designed to calibrate all the different lines with the help of the variable antenna coupling.
By moving the antenna one can arbitrarily choose the coupling to a mode, if it is weakly coupled a test signal from the Aux port gets reflected and goes to the JPA, while if the antenna is critically coupled to the mode, a signal from SO is transmitted through the cavity and than to the JPA. Almost the same result can be obtained by slightly changing the frequency of the test signal to be within the JPA band but out of the cavity resonance. The critical coupling can be reached by doubling the linewidth of the mode or equivalently by minimizing the reflected signal from the Aux line to the Readout line. The procedure to calibrate all the lines is:
\begin{enumerate}
\item with the weakly coupled antenna or by detuning the mode the losses of the Aux-Readout line $L_\mathrm{AR}$ are measured;
\item the antenna is critically coupled to the mode and a signal is sent through the Aux-SO line to get $L_\mathrm{AS}$;
\item with the same critical coupling the transmission of the SO-Readot line $L_\mathrm{SR}$ is acquired.
\end{enumerate}
At this point a signal of power $A_\mathrm{in}$ is injected in the SO line, the fraction of this power getting into the cavity through the weakly coupled antenna is $A_\mathrm{cal}=A_\mathrm{in}L_\mathrm{SO}$.
The attenuation of the line can be calculated as $L_\mathrm{SO}\simeq\sqrt{L_\mathrm{SR}L_\mathrm{AS}/L_\mathrm{AR}}$, which gives the power collected by the critically coupled antenna. Since $A_\mathrm{cal}$ is effectively a calibrated signal, it can be used to measure gain and noise temperature of the Readout line. Different $A_\mathrm{in}$ are used to get increasingly large signals to be detected by the JPA-based chain.
This calibration has some minor biases, the first is given by the cable from the cavity to the first circulator which is accounted for two times in the Aux-Readout line. This contribution can be safely neglected as the cable is superconducting, making its losses negligible.
Another bias is related to the antenna coupling, which is not perfect. With a proper antenna coupling the reflected signal is reduced of $\sim10$\,dB, so there is a bias of a factor 10\% intrinsic to the measurement which will be accounted for when calculating the error.
As the calibration procedure is long it is not repeated for every run, however no important differences are expected when changing the JPA frequency.
As an example a run at 10.409\,GHz is considered. The gain of the JPA at this frequency results $G_\mathrm{JPA}\simeq18\,$dB, and its bandwidth is 8\,MHz, the hybrid mode is tuned the its central frequency and the calibration procedure is carried out,resulting in a noise temperature of $T_n=1.0\,$K and the total gain of the whole amplification chain is $G_\mathrm{tot}\simeq120.4\,$dB. The value of $T_n$ is compatible with the one estimated previously, and corresponds to two quanta. 

The coupling of the antenna with the hybrid mode is checked for every run. It is controlled by moving the dipole antenna in and out the cavity volume, the critical coupling is reached when the uncoupled linewidth of the mode is doubled.
To verify the proper antenna positioning one may rely on the fact that depending on the temperature difference between the cavity and a 50\,$\Omega$, some power may be absorbed or added to the load thermal noise.
The load under consideration is the hottest between the first JPC isolator and the 20\,dB attenuator of the Aux line. The hybrid resonance has a critical linewidth of about 2\,MHz, so the depth will not be as narrow as the one of the cavity. In that case the temperature difference is about 10\%, which is about 10\,mK, and if the temperature of the load and cavity are precisely measured the spectra can be used to get a two-points calibration. The selected calibration procedure was not this one because the temperatures of loads and HS are not easily accessible. With two dedicated thermometers the temperatures of the loads could be measured, but it is not trivial to measure the temperature of the cavity and of the spheres with the needed precision. Since a small temperature difference is expected, the measurement with the antenna coupled to the hybrid mode should be different from the uncoupled one. As reported in \cite{PhysRevLett.124.171801}, there is a difference between the two measurements and it is compatible with the thermal noise of the hybrid mode at a temperature slightly higher then the loads one.

\section{Data acquisition and analysis}
\label{sec:daq}
A ferromagnetic haloscope's scientific run consists in several measurements in the common bands between the frequencies of the lower hybrid mode unaffected by disturbances, and the JPA working range. The low temperature electronics is described in the previous section, and is completed by its following part hereafter.

The room temperature electronics consists of a HEMT amplifier (A2) followed by an IQ mixer used to down-convert the signal with a local oscillator (LO).
In principle, it is possible to acquire the signal coming from both hybrid modes using two mixers working at $f_+$ and $f_-$. In this case it is chosen to work only with $f_+$, thus setting the LO frequency to $f_\mathrm{LO}=f_+-0.5\,$MHz. The amplified antenna output at the hybrid mode frequency is down-converted in the 0 - 1\,MHz band, allowing to efficiently digitize the signal.
The phase and quadrature outputs are fed to two low frequency amplifiers (A3$_{I,Q}$), with a gain of $G_3\simeq50\,$dB each, and are acquired by a 16 bit ADC sampling at 2\,MS/s (see \cite{quaxepjc}).
A dedicated DAQ software is used to control the oscillators and the ADC, and verifies the correct positioning of the LO with an automated measurement of the hybrid mode transmission spectrum. Some other online checks include a threshold monitor of the average amplitude, as well as of the peak amplitude, which flags the file if some unexpected large signal is present. 
The ADC digitizes the time-amplitude down-converted signal coming from A3$_I$ and A3$_Q$ and the DAQ software stores collected data binary files of 5\,s each.
The software also provides a simple online diagnostic, extracting 1\,ms of data every 5\,s, and showing its 512 bin FFT together with the moving average of all FFTs.
The signal is down-converted in its in-phase and quadrature components $\{\phi_n\}$ and $\{q_n\}$, with respect to the local oscillator, that are sampled separately.

	\begin{figure}[h!]
	\centering
	\resizebox{0.6\hsize}{!}{ \includegraphics*{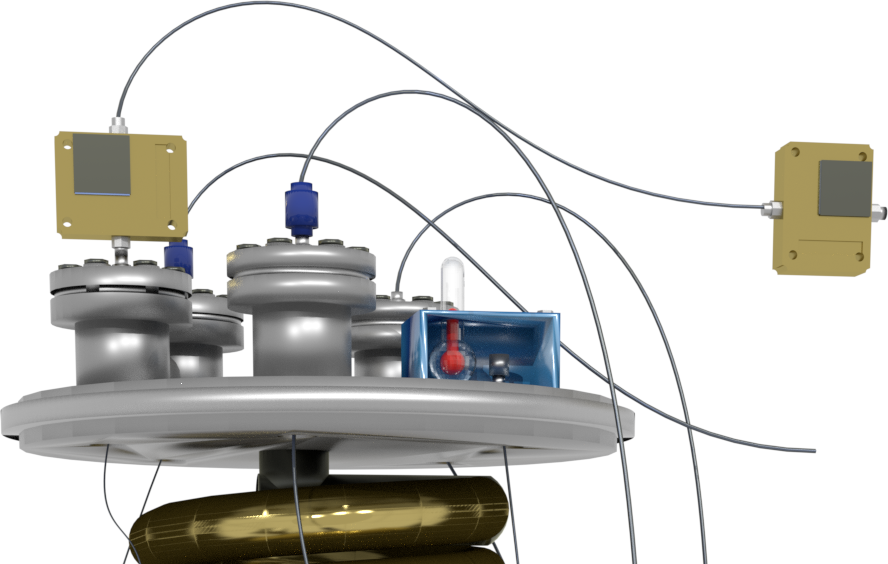}} 
	\caption{Second amplification stage of the setup, and first room temperature amplifier. The image shows the top part of the vacuum vessel containing the dilution fridge stages, the cavity and the electronics. Just outside it, the first amplifier is A1 (reported in yellow), while the second one is already at room temperature. The blue box corresponds to the variable temperature load of configuration (b).}
	\label{fig:daq}
	\end{figure}

The stability of the measurement is tested by injecting a signal in the SO line slightly off resonance with the PMHS peak, and with an amplitude guaranteeing a large SNR. Monitoring its amplitude is a way to continuously check the peak position. In this setup the stability results well below the percent level, which is more than enough for the purpose of the experiment, thanks to the lower and more stable working temperature and to an extremely stable current generator produced in the Padua University electronic workshop.

The signal is analysed using a complex FFT on the combination of phase and quadrature $\{s_n\}=\{\phi_n\}+i\{q_n\}$ to get its power spectrum $s^2_\omega$ with positive frequencies for $f>f_\mathrm{LO}$ and negative frequencies for $f<f_\mathrm{LO}$, in a total bandwidth of 2\,MHz which contains the whole hybrid linewidth. Some bins are found to be affected by disturbances resulting in systematic noise, but they can be easily removed by analyzing the data with longer FFT. Using 32768 points the single frequency resolution is $61\,\mathrm{Hz}\ll \gamma_a$ and thus single bins which do not respect the fluctuation dissipation theorem can be substituted with the average of the ten nearest neighbors. This procedure artificially reduces the variance of the data, but the number of corrupted bins is negligible and so is the effect on the variance. This process does not cut the signal since it affects only known or single bins, while the axion signal is expected to be distributed over many.

For debugging purposes a simulated signal is injected into the analysis code. It is created by generating a high-frequency noise to which are added a simulated axion signal and some disturbances. For creating the in-phase and in-quadrature component it is multiplied by a sine or to a cosine wave, to then extract only one point every $10^4$ and simulate the mixing and down-conversion processes. The FFT of this signal produces a white noise with some peaks (the axion signal plus disturbances), and is eventually integrated. The analysis procedure is verified to remove bad bins and to preserve the signal and SNR.

Thanks to the stability and to the tests on the acquisition and analysis, all the files of a single run can be safely RMS averaged together.
The calibration gives a noise temperature $T_n=1.0\,$K, which includes all the amplifier noises, the cavity thermodynamic noise, and the losses. This can be used to calibrate the setup by setting the mean value of the FFT to $k_B T_n \Delta f$.
As for the calibration errors, the fluctuations of the cavity temperature are of order 10\,mK, thus they can change $T_n$ of a fraction close to 1\%. A larger contribution is intrinsic to the procedure, which requires the coupling of the antenna to be changed from weak to critical.
This is believed to be the larger contribution to the uncertainty of the measurement, and even if in principle this is not a fluctuation it will be used to estimate an error. The fact that the coupling is close to critical is also supported by a thermal noise measurement \cite{PhysRevLett.124.171801}.
The control over this parameter can be estimated by injecting a signal through the Aux port in the weak-coupling position, and then by reducing the reflected amplitude by increasing the coupling. Typically the signal power can be decreased of more than 8\,dB, this value can be used to estimate the larger uncertainty on the coupling resulting in a 16\% error.

To estimate the sensitivity to the axion field the resolution bandwidth is set to 5\,kHz, which at this frequency is the value producing the best SNR. The spectrum is fitted with a degree five polynomial to extract the residuals, whose standard deviation is the sensitivity of the apparatus in terms of power. For every run, by using the run length, the RMS power is compared with the estimated sensitivity obtained with Dicke radiometer equation.
The measured fluctuations are close to the expectations for almost every run, and are always compatible when considering the 16\% error which was previously calculated.
This shows that the measured power is compatible thermal noise and that the integration is effective over the whole measurement time, since it follows the $1/\sqrt{\mathrm{time}}$ trend.
The best sensitivity reached is $\sigma_P=5.1\times10^{-24}\,$W for an integration time of 9.3\,h, corresponding to an excess rate $\lesssim10$\,photons/s per bin.

As is discussed in Section \ref{intro} this setup is a spin-magnetometer, as it is sensitive to variations of the sample magnetization.
In this sense, it is interesting to determine the performances of this setup to get some physical intuition on its sensitivity and thus on all the possible phenomena that could be measured.
The field sensitivity can be estimated for a field whose coherence length stretches on all the YIG spheres, and whose linewidth is narrower than the PMHS one. In the case of the present apparatus the threshold coherence length and time are then 10\,cm and 100\,ns, respectively.
If $f_-$ and $\tau_-$ are the frequency and coherence time of the hybrid mode and $N_S$ is the number of spins, the magnetic field sensitivity of the setup is
    \begin{equation}
    \sigma_B=\,\Big( \frac{\sigma_P}{4\pi \gamma_e \mu_B f_- \tau_-  N_S}\Big)^{1/2}
                         =\,5.5\times10^{-19} \Big[\Big( \frac{10.4\,\mathrm{GHz}}{f_-} \Big) \Big( \frac{83\,\mathrm{ns}}{\tau_-} \Big) \Big( \frac{1.0\times10^{21}\,\mathrm{spins}}{N_S} \Big)\Big]^{1/2}\,\mathrm{T},
    \label{field_limit}
    \end{equation}
which is the minimum effective magnetic field detectable by the spin magnetometer for a unitary SNR.
A 0.5\,aT sensitivity is remarkable by itself, and shows the high potential of HS-based magnetometers.

The results of the previous section are used to extract a limit on the axion-electron coupling, which is a possible effect that modulates the sample magnetization and produces an excess of photons.
The expected power deposited by DM-axions in the PMHS is
    \begin{equation}
    P_{\rm out}=7 \times 10^{-33} \left(\frac{m_a}{43\, \mu{\rm eV}}\right)^3 \left( \frac{N_S}{1.0\times10^{21}\,\mathrm{spins}}\right)  \left( \frac{\tau_h}{83\,{\rm ns}}\right)\,{\rm W}.
    \label{eq:4_pin2019}
    \end{equation}
The 10.4\,GHz photon rate corresponding to this power is $r_a=10^{-9}\,$Hz.
By comparing the rate $r_a$ with $\sigma_P^\mathrm{(3)}/(\hbar\omega_-)$ one obtains that there are ten orders of magnitude to get the sensitivity required to detect the axion, which corresponds to five in terms of field (and thus of coupling constant).
As discussed in Section \ref{intro}, even if an instrument is not capable of limiting the QCD-axion parameter space it can still probe the presence of ALPs. These can also constitute the totality of DM and their mass and coupling are unrelated.
The present measurements are used to extract a limit on the coupling of ALPs with electrons reported in \cite{PhysRevLett.124.171801}. No evidence of a signal due to axions has been detected, and the measured spectra are compatible with noise. The $2\sigma$ (95\% C.\,L.) upper limit on the coupling reads
    \begin{equation}
    g_{aee}>\frac{e}{\pi m_av_a}\sqrt{\frac{2t_{ac}\times\sigma^\mathrm{(3)}_P}{2\mu_B \gamma\, n_a N_S \tau_-}},
    \label{eq:4_gaee3}
    \end{equation}
where $t_{ac}$ is a frequency dependent coefficient that takes into account that the axion-deposited power is not uniform in the haloscope operation range, as is discussed in \cite{Crescini2020}.
All the experimental parameters used to extract the limits are measured within every run, making the measurement highly self-consistent.
The limit on the ALP-electron coupling described by Eq.\,\ref{eq:4_gaee3} is calculated for every run using the corresponding measured parameters. This result is compared with other techniques used for testing the axion-electron couling constant 
    \begin{figure}[h!]
    \centering
    \resizebox{0.6\hsize}{!}{ \includegraphics*{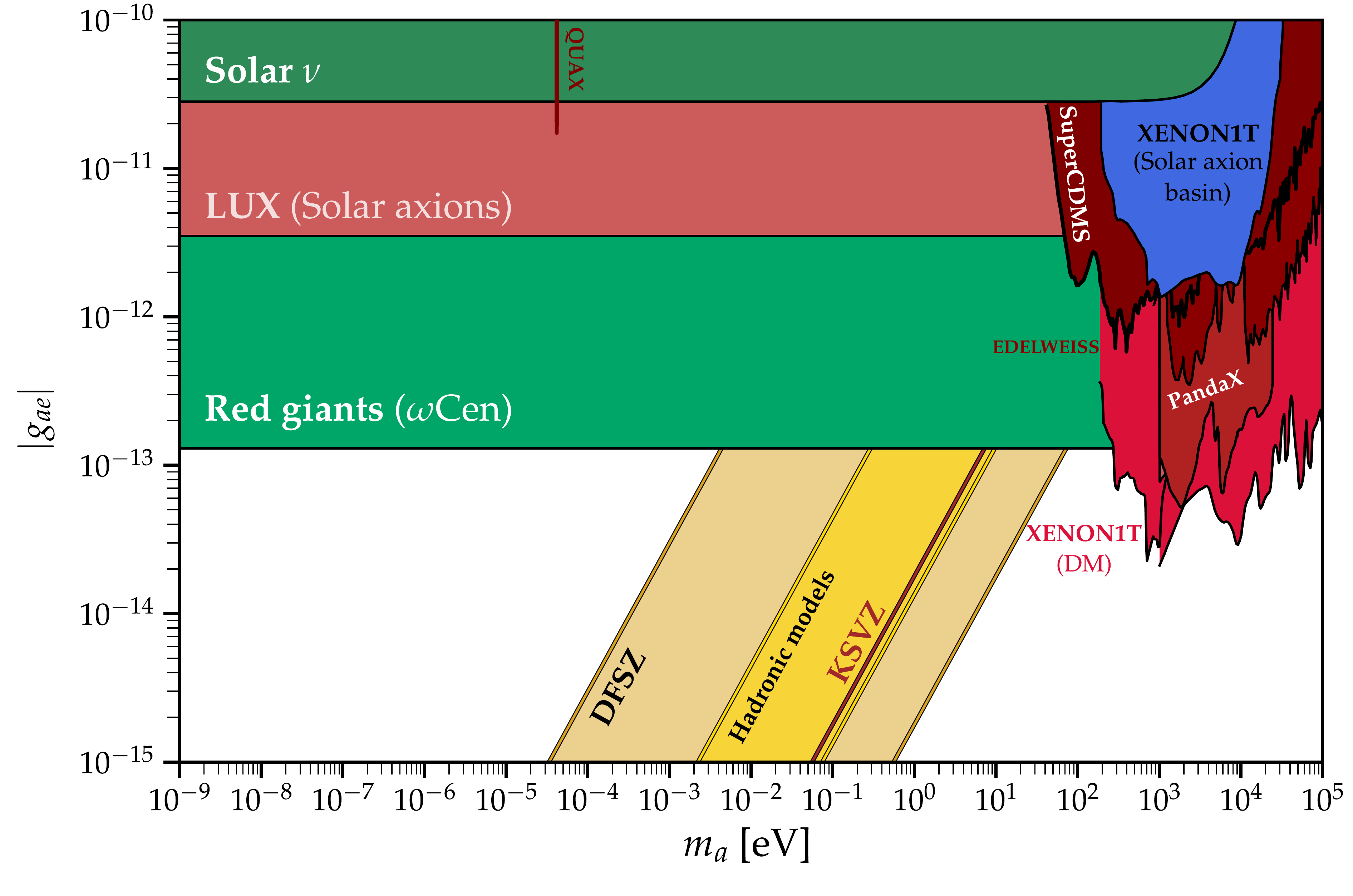}} 
    \caption{Overview of axion searches based on their coupling with electrons \cite{ciaran_o_hare_2020_3932430}. The result obtained with the ferromagnetic haloscpe described in this work is labelled as ``QUAX''.}
    \label{fig:gaee}
    \end{figure}
    
The analysis is repeated by shifting the bins of half the RBW to exclude the possibility of a signal divided into two bins. The best limit obtained, and corresponding to $\sigma_P$, is $g_{aee}<1.7\times10^{-11}$.
The improvement of a longer integration time is not much, since the limit on the coupling scales as the fourth root of time, and to improve the current best limit of a factor 2 the needed integration time is six days.

\section{Conclusions}
\label{sec:conclusions}
\begin{figure}[h!]
    \centering
    \resizebox{0.6\hsize}{!}{ \includegraphics*{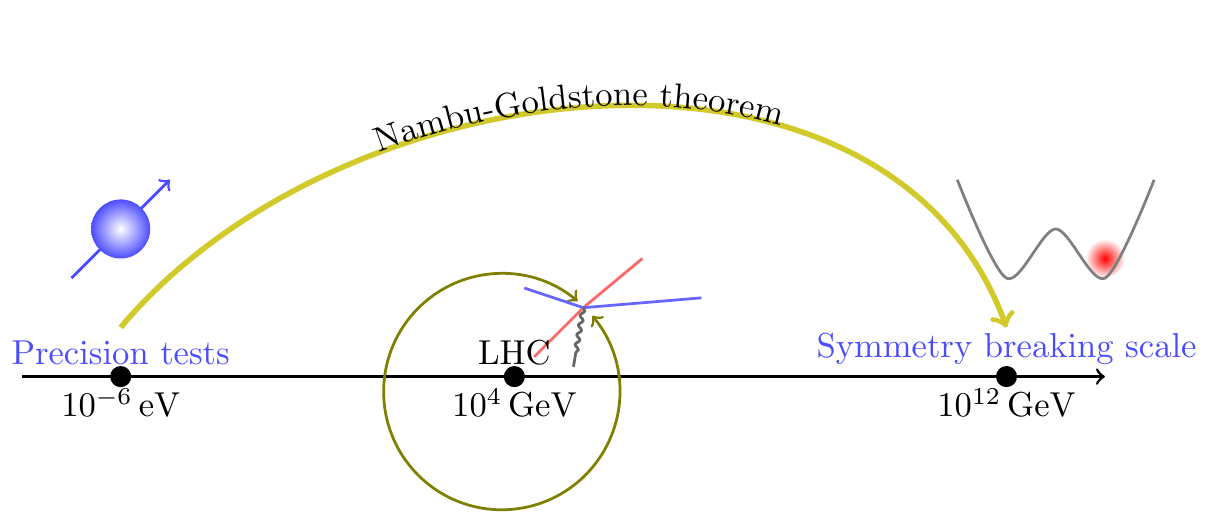}} 
    \caption{The usage of Nambu-Goldstone theorem to infer on physics at energy scales inaccessible to accelerators. }
    \label{fig:nambu}
    \end{figure}

Low-energy measurements, precisely testing known physical laws, are a powerful probe of BSM physics, mainly complementary to accelerator physics. As shown in Fig.\,\ref{fig:nambu}, thanks to Nambu-Goldstone theorem, extremely high energy scales can be explored by measuring faint effects at the limit of present technology. Eventually, new instruments and devices can be built to push the current technological limits to new levels and hopefully help not only fundamental physics but many other fields.

Among these, haloscopes play a pivotal role while searching for Dark Matter axions.
The scope of this work is to illustrate the construction and outline the operation of the first ferromagnetic axion haloscopes.
Such instruments can be used to measure the DM-axion wind which blows on Earth, as this last one is moving through the halo of the Milky Way. The axions interact with the spin of electrons causing spin flips that are, macroscopically, oscillations of a sample magnetization. The model of the isothermal galactic halo and of the axion yield the features of the searched for signal, namely its linewidth, frequency and amplitude. A frequency of $10\,$GHz, a linewidth of $5\,$kHz and an amplitude of about $10^{-23}\,$T are expected for axion masses of order 40$\,\mu$eV.
A proper haloscope has a transducer that converts the axion flux into rf power, followed by a sensitive detector to measure it.
For a ferromagnetic haloscope, the transducer consists of a magnetic material containing the electron spins with which the axions interact.
In order to maximize the axion-deposited power, the sample should have a large spin density and a narrow linewidth. The detector is a rf amplification chain based on a JPA.
The described instrument features a power sensitivity limited by quantum fluctuations, in this sense no linear amplifier is or will be able to improve the haloscope. In future setups only bolometers or quantum counters can yield better results. For example, recent developments on quantum technologies \cite{PhysRevX.10.021038,Albertinale2021} demonstrated the detection of fluorescence photons emitted by an electron spin ensemble, and could be adopted for axion searches.

Thermodynamic fluctuations are already negligible due to the extremely low working temperature, so it is not necessary to decrease them by orders of magnitude. As the rate of thermal photons of a cavity mode is exponentially decreasing with temperature, the present dilution refrigeration technology is enough to reach the axion-required noise level.
As discussed in Section \ref{intro}, to get a rate of axion-induced photons which can be measured in a reasonable amount of time, a much increased quantity of material and a narrower linewidth are required. This setup features 0.05\,cc of YIG, such volume must be increased by three orders of magnitude to get the required rate. This large quantity can be achieved by increasing the quantity of material in a single cavity and the number of cavities.

In conclusion, the successful operation of an ultra cryogenic quantum-limited prototype demonstrates the possibility of scaling up the setup of orders of magnitude without compromising its sensitivity. To further increase the axionic signal there are two parameters to work on: the hybrid mode linewidth and the sample spin-density and volume. To finally achieve the sensitivity required by a QCD axion search, it is necessary to use a photon counter.
The upgrades planned until now are implemented and result effective, as the apparatus behaves as expected. No showstoppers were identified so far.
 
\section{Acknowledgment}
N.C. is thankful INFN and the Laboratori Nazionali di Legnaro for hosting and encouraging the experiment. 
The help and support of Giovanni Carugno and Giuseppe Ruoso is deeply acknowledged.

 \bibliographystyle{unsrt_3}
\bibliography{buildingHaloscopes}

\end{document}